\documentclass[a4paper]{article}
\usepackage[psamsfonts]{amssymb}
\usepackage{amsmath}
\usepackage{cite}
\usepackage{epsfig}
\usepackage{graphicx}
\date{}

\author{M. Alimohammadi\footnote{alimohmd@ut.ac.ir}\ \ and N. Agharafiei
\\ {\small Department of Physics, University of Tehran,}
\\ {\small North Karegar Ave., Tehran, Iran.}}
\title{Quantum attractors of generalized Gauss-Bonnet dark energy }
\begin{document}
\maketitle
\begin{abstract}
The influences of quantum effects on the structure of the
phase-space of generalized Gauss-Bonnet theory, introduced by the
Lagrangian $F(R,G)$, have been studied. $G$ is the Gauss-Bonnet
invariant, and the quantum effects are described via the account
of conformal anomaly. It has been shown that the quantum effects
change many aspects of the attractors of $F(R,G)$ gravity models
in the $R-H$ plane, including the location of the attractors, the
number of them and their stability properties. These variations
are not, in general, from the type of small perturbations, but
instead, it can induce the great, so not ignorable, variations
which have root in the "singular perturbation" nature of this
effect. In other words, one can not ignore the quantum corrections
and must be always considered. The influences of the perfect
barotropic fluids on this problem have been studied, and it has
been shown that this kind of matters do not alter the quantum
effects. It has been shown that the classical contribution of the
coupled-quintessence model, which is responsible for inducing the
quantum effects, is of this type, that is a perfect barotropic
fluid, and therefore can not change our results.
\end{abstract}
\section{Introduction}
Independent observational data indicate that we are now in
accelerating phase of the universe. The supernova observations
directly support this accelerating expansion, and the microwave
background, large scale structure and its dynamics, weak lensing
and baryon oscillation observations indirectly verify this
phase~\cite{perl}. It is believed that nearly 70$\%$ of our
present universe is composed of dark energy, the physical object
which is responsible for the effective negative pressure required
for accelerating cosmic expansion.

There are two main dynamical models that explain some features of
dark energy. The first one is based on the standard Einstein
theory, when some other physical terms added to it. The simple
cosmological constant model, with no dynamics~\cite{wein}, the
scalar field models, including quintessence, phantom, quintom and
hessence models~\cite{wett}, and some other models like the
k-essence models, chaplygin gas models, and holographic dark
energy models~\cite{chib} are examples of the first class of dark
energy models.

The second class of dark energy models are those based on the
assumption that the gravity is being (nowadays) modified. These
modified gravity theories can be divided to two main subclasses.
The first subclass consists of the models in which, besides the
scalar-curvature $R$, there exists a scalar field (or either a
vector field) which interacts nonminimally to gravity. These
models are known as the scalar-tensor models, where after the
simplest Brans-Dicke theory~\cite{jord}, several generalizations
have been introduced~\cite{berg}. Recently it has been shown that
the generalized scalar-tensor theories admit the $\omega=-1$ and
deceleration to acceleration transitions, the phenomenon which can
be affected by quantum effects~\cite{alb}. The equation of state
parameter $\omega$ is defined through $\omega=p/\rho$, where $p$
is the pressure and $\rho$ is the energy density of dark energy.
In cosmological constant model, $\omega$ takes the constant value
$\omega_{\Lambda}=-1$, and in dynamical models, it is a function
of time, i.e. $\omega=\omega(t)$.

The second subclass of modified gravity models are those whose
actions are not the simple term $R$, but instead they are, in
general, an arbitrary function of all algebraic invariants built
up with the Riemann tensor, including the scalar curvature $R$,
the quadratic invariants  $P=R^{\mu\nu}R_{\mu\nu}$ and
$Q=R^{\alpha\beta\gamma\delta}R_{\alpha\beta\gamma\delta}$, and
other independent invariants of higher orders. The $f(R)$ gravity,
whose Lagrangian is an arbitrary function $f(R)$, is the simplest
and the most famous modified gravities of this subclass, i.e.
\begin{equation}\label{1}
S=\int d^{4}x\sqrt{-g}{}\hspace{1ex}\left[
\frac{1}{2\kappa^{2}}f(R)+{\cal L}_m \right].
\end{equation}
In $\hbar=c=G=1$ units, $\kappa^{2}=8\pi$ and ${\cal L}_m$ is the
Lagrangian density of dust-like matter. Many aspects of $f(R)$
gravity, including their local properties, have been studied.
See~\cite{noji} and references therein.

Restricting ourselves to quadratic invariants, the generalized
modified gravity models are defined through
\begin{equation}\label{2}
S=\int d^{4}x\sqrt{-g}{}\hspace{1ex}\left[f(R,P,Q)+{\cal L}_m
\right],
\end{equation}
where $f(R,P,Q)$ is an arbitrary function. Some aspects of these
models, including their attractor solutions and their stabilities
have been discussed in~\cite{carr}. Studying the propagators of
these models has shown an important result, i.e. the generalized
modified gravity models have, in general, the graviton ghost,
unless it satisfies~\cite{nune}
\begin{equation}\label{3}
\left(f_{P}+4f_{Q}\right)|_{R=R_{0}}=0,
\end{equation}
where the subscripts denote the partial derivatives, e.g.
\begin{equation}\label{4}
f_{P}=\frac{\partial f(R,P,Q)}{\partial P},
\end{equation}
and the curvature $R_{0}$ is defined by the equation
\begin{equation}\label{5}
f-\frac{1}{2}R_{0}f_{R}-\frac{1}{4}{R_{0}}^{2}f_{P}-\frac{1}{6}{R_{0}}^{2}f_{Q}=0.
\end{equation}
The above relation is the equation of motion of the metric at
constant scalar curvature $R=R_{0}$. In eq.(\ref{5}),
$Q={{R_{0}}^{2}}/{4}$ and $P={{R_{0}}^{2}}/{6}$. Now if one
considers the Gauss-Bonnet (GB) invariant $G$
\begin{equation}\label{6}
G=R^{2}-4P+Q=R^{2}-4R_{\mu\nu}R^{\mu\nu}+R_{\mu\nu\xi\sigma}R^{\mu\nu\xi\sigma}.
\end{equation}
it is easily seen that the condition (\ref{3}) satisfies, and this
is a reason why the generalized GB gravity, defined by
\begin{equation}\label{7}
S=\int d^{4}x\sqrt{-g}{}\hspace{1ex}\left[F(R,G)+{\cal
L}_m\right],
\end{equation}
is an important candidate of modified gravity theories.

The generalized GB gravity, where the $f(R)$ gravity (\ref{1}) is
a special example of it, has been first introduced in~\cite{cogn}.
The hierarchy problem of particle physics and the late time
cosmology in the context of $F(R,G)$, have been studied
in~\cite{cogn} and~\cite{km}, and its behavior under
phantom-divide-line crossing and deceleration to acceleration
transitions has been investigated in~\cite{alai}. Recently, the
two-dimensional phase-space of the generalized GB models, i.e. the
$R-H$ space where $H$ is the Hubble parameter, has been studied
in~\cite{alaii} and the various aspects of the de-Sitter
attractors of these models have been discussed.

The present paper is devoted to study the contribution of quantum
effects to the attractors of generalized GB gravity. The quantum
effects are described via the account of conformal anomaly,
reminding about anomaly-driven inflation~\cite{star}. The
contribution of conformal anomaly in energy conditions and big rip
of phantom models has been discussed in~\cite{nojii}, and its
influence on the $\omega=-1$ crossing and deceleration to
acceleration transition of quintessence and phantom models, the
$F(R,G)$ gravity and the generalized scalar tensor models have
been discussed in~\cite{als},~\cite{alai} and~\cite{alb},
respectively.

The late-time behavior of dynamical models, including the dark
energy models, is an important problem, both in mathematics and
physics, and is studied in a framework known as the attractor
solutions of dynamical systems. Many properties of attractor
solutions of dark energy models have been studied~\cite{cope}. Two
important parts of attractor studies of any dynamical system are:
a) choosing the phase-space of the model and b) obtaining the
location of attractors and specifying the stability of them. In
$F(R,G)$ gravity models, the only possible choice of phase-space
for general $F(R,G)$ model is $R-H$ space. So attractors, with
properties $\dot R=0$ and $\dot H=0$, or $R=R_{c}$ and $H=H_{c}$
($R_{c}$ and $H_{c}$ are constant values ), are de-Sitter
solutions of generalized GB models. Now, as we will show, the
quantum effects have important, and non-ignorable, contributions
to this problem.  It changes the position of the attractors and
their stability behaviors, and, for some cases, it produces the
new attractors where some of them do not lead to classical
attractors, a phenomena known as "singular perturbation" in
mathematics. This shows that the quantum effects are not
ignorable, i.e. they do not only lead to small corrections for
classical solutions. For some other cases, it can remove the
degeneracy of solutions, which is known as "bifurcation" in the
mathematics of dynamical systems. And, interestingly, it can
transform the critical curves, an infinite number of stable
attractors locating on $R=12H^{2}$ curve in the phase-space of
some specific case of $F(R,G)$ models, to a unique point.

The scheme of the paper is as follows: In section 2, we discuss
the critical points and the stability conditions of $F(R,G)$
models in the presence of quantum correction terms. It is shown
that in the $\hbar\rightarrow 0$ limit, the relations are
correctly reduced to classical relations. In section 3, we discuss
the various examples of $f(R)$ and $F(R,G)$ models. The new
aspects of the phase-space of these models, including those
introduced in the last paragraph, can be seen via these examples.
It is shown that the numerical calculations verify our results.
The usual matter's contributions to the location and behavior of
attractors are discussed in section 4. It is shown that the
barotropic fluids can not change the structure of the phase-space
of $F(R,G)$ models. It is also shown that the classical
contributions of the coupled-quintessence model, which is
responsible for producing the quantum correction terms considered
in section 2, can not alter our results, which is an important
consequence.

\section{Quantum corrected critical points of $F(R,G)$ gravity}
Consider the generalized GB dark energy with action (\ref{7}).
Variation of this action with respect to the metric $g_{\mu\nu}$
results in~\cite{cogn}
\begin{equation}\begin{split}\label{8}
&\frac{1}{2}T^{\mu\nu}
+\frac{1}{2}g^{\mu\nu}F(R,G)-2F_{G}(R,G)RR^{\mu\nu}
+4F_{G}(R,G)R^{\mu}_{~\rho}R^{\nu\rho}\\
&-2F_{G}(R,G)R^{\mu\rho\sigma\tau}R^{\nu}_{~\rho\sigma\tau}
-4F_{G}(R,G)R^{\mu\rho\sigma\nu}R_{\rho\sigma}
+2(\nabla^{\mu}\nabla^{\nu}F_{G}(R,G))
R\\&-2g^{\mu\nu}(\nabla^{2}F_{G}(R,G))
R-4(\nabla_{\rho}\nabla^{\mu}F_{G}(R,G))
R^{\nu\rho}-4(\nabla_{\rho}\nabla^{\nu}F_{G}(R,G))R^{\mu\rho}\\
&+4(\nabla^{2}F_{G}(R,G))R^{\mu\nu}
+4g^{\mu\nu}(\nabla_{\rho}\nabla_{\sigma}F_{G}(R,G))R^{\rho\sigma}
-4(\nabla_{\rho}\nabla_{\sigma}F_{G}(R,G))R^{\mu\rho\nu\sigma}\\
&-F_{R}(R,G)R^{\mu\nu}+\nabla^{\mu}\nabla^{\nu}F_{R}(R,G)-
g^{\mu\nu}\nabla^{2}F_{R}(R,G)=0 .
\end{split}\end{equation}
$T_{\mu\nu}$ is the energy-momentum tensor, defined through
\begin{equation}\label{9}
T_{\mu\nu}=\frac{2}{\sqrt{-g}}\frac{\delta S_{m}}{\delta
g_{\mu\nu}},
\end{equation}
and the subscripts of $F(R,G)$ denote the partial derivatives. For
the background metric, we consider the flat
Friedmann-Robertson-Walker (FRW) metric, defined by
\begin{equation}\label{10}
ds^{2}=-dt^{2}+a^2(t)(dx^{2}+dy^{2}+dz^{2}).
\end{equation}
$(t,x,y,z)$  are comoving coordinates and $a(t)$ is the scale
factor. The three independent Friedmann equations of $F(R,G)$
gravity, then become~\cite{alai}
\begin{align} \label{11}
 -6H^{2}F_{R}(R,G)&=F(R,G)-RF_{R}(R,G)+6H\dot{F}_{R}(R,G)
 \nonumber \\
                  &+24H^{3}\dot{F}_{G}(R,G)-GF_{G}(R,G) -\rho_{m},
\end{align}
\begin{equation}\label{12}
R=6(\dot H+2H^{2}),
\end{equation}
\begin{equation}\label{13}
G=24H^{2}(\dot H+H^{2}),
\end{equation}
and the evolution equation of matter field is
\begin{equation}\label{14}
\dot{\rho}_{m}+3H(\rho_{m}+p_{m})=0 .
\end{equation}
$H=\dot{a}(t)/a(t)$  and "dot" denotes the time derivative.

To study the quantum effects, we consider the following standard
Lagrangian of a scalar field $\phi$, known as the
coupled-quintessence model, in which $\phi$ is nonminimally
coupled to gravity
\begin{equation}\label{15}
{\cal{L}}=\frac{1}{2}\left(-\xi R\phi^2-(\nabla\phi)^2\right),
\end{equation}
where the renormalizibility requirements force $\xi$ to be
$\xi={1}/{6}$. Calculating the effective action of this conformal
invariant Lagrangian at one-loop level, results in a nonvanishing
trace for the energy-momentum tensor, which is classically
traceless. This trace, i.e. the trace/conformal anomaly,
is~\cite{star,birrl}
\begin{equation}\label{16}
T_{A}=b(F+\frac{2}{3}\square R)+b'G+b''\square R\hspace{1ex}.
\end{equation}
The subscript "$A$" denotes "anomalous", $G$ and $R$ are
Gauss-Bonnet and Ricci scalars, and $F$ is the square of the 4$d$
Weyl tensor
\begin{equation}\label{n17}
F=\frac{1}{3}R^{2}-2R_{\mu\nu}R^{\mu\nu}+R_{\mu\nu\alpha\beta}R^{\mu\nu\alpha\beta}\hspace{1ex}.
\end{equation}
$b$, $b'$ and $b''$ are given by
\begin{equation}\begin{split}\label{17}
&b=\frac{3(N+6N_{1/2}+12N_{1}+611N_{2}-8N_{\rm{HD}})}{360(4\pi)^{2}}\hspace{1ex} ,\\
&b'=-\frac{N+11N_{1/2}+62N_{1}+1411N_{2}-28N_{\rm{HD}}}{360(4\pi)^{2}}
,\hspace{1ex}b''=0.
\end{split}\end{equation}
Eq.(\ref{17}) is for the cases where there exist $N$  scalars,
$N_{1/2}$ spinors, $N_{1}$ vector fields and $N_{2}( = 0, 1)$
gravitons, and $N_{\rm{HD}}$ higher derivative conformal scalars.
The energy density $\rho_{A}$ and pressure $p_{A}$, corresponding
to $T_{A}$, can be found by:
\begin{equation}\label{18}
\rho_{A}=-\frac{1}{a^4}\int_{0}^{t}a^{4}HT_{A}\,dt,
\end{equation}
and
\begin{equation}\label{19}
p_{A}=\frac{1}{3}\left(T_{A}+\rho_{A}\right),
\end{equation}
respectively. The resulting relations in FRW metric
are~\cite{nojii}
\begin{equation}\begin{split}\label{20}
&\rho_{A}=-6b'H^{4}-(\frac{2}{3}b+b'')(-6H\ddot{H}-18H^{2}\dot{H}+3{\dot{H}^{2}})\hspace{1ex} ,\\
&p_{A}=b'(6H^{4}+8H^{2}\dot{H})+(\frac{2}{3}b+b'')(-2\dddot{H}-12H\ddot{H}-18H^2\dot{H}-9{\dot{H}}^{2}).
\end{split}\end{equation}
Note that in SI units, the above relations have $\hbar/c^{3}$ as
the prefactor. It can be easily shown that the above expressions
satisfy
\begin{equation}\label{21}
{\dot\rho}_{A}+3H(p_{A}+\rho_{A})=0.
\end{equation}
The natural method for computing the quantum corrections in
$F(R,G)$ gravity models is to add $\rho_{A}$ to Friedmann
equations, i.e. to change $\rho_{m}$ in eq.(\ref{11}) to
$\rho_{m}+\rho_{A}$.

Using eqs.(\ref{12}) and (\ref{13}), one finds
\begin{equation}\label{22}
G=4H^{2}(R-6H^{2}),
\end{equation}
from which
\begin{equation}\label{23}
\dot{G}=\frac{4}{3}HR^{2}+192H^{5}-32RH^{3}+4H^{2}\dot{R}.
\end{equation}
Also eq.(\ref{20}) leads to
\begin{equation}\label{24}
\rho_{A}=-6b'H^{4}+\left(\frac{2}{3}b+b''\right)\left(H\dot{R}+RH^{2}-\frac{R^{2}}{12}\right).
\end{equation}
The time derivatives in eq.(\ref{11}) can be also written in terms
of $\dot{R}$ and $\dot{G}$, using
\begin{equation}\label{25}
\frac{d}{dt}f(R,G)=f_{R}\dot{R}+f_{G}\dot{G}\;.
\end{equation}
In this way, $\dot{R}$ and $\dot{H}$ are found from Friedmann
equations as following
\begin{equation}\begin{split}\label{26}
\dot{R}=&\frac{1}{6H(F_{RR}+8H^{2}F_{RG}+16H^{4}F_{GG})-(\frac{2}{3}b+b'')H}
\times \bigg\{ (R-6H^{2})F_{R}+GF_{G} \\ &
-288H^{2}{(\frac{R}{6}-2H^{2})}^{2}(F_{RG}+4H^{2}F_{GG}) +
(\frac{2}{3}b+b'')(RH^{2}-\frac{R^{2}}{12})-6b'H^{4}+\rho_{m}
\bigg\},
\end{split}\end{equation}
\begin{equation}\label{27}
\dot{H}=\frac{R}{6}-2H^{2}.
\end{equation}
These are the set of autonomous equations, where its phase-space
is two dimensional $(R-H)$ space. Here, for simplicity, we do not
consider the matter field, i.e. $\rho_{m}=0$. We will come back to
matter field in section 4. Note that in the right-hand-side of
eq.(\ref{26}), the relation (\ref{22}) must be used for $G$.

Our autonomous equations are in the form
\begin{equation}\begin{split}\label{28}
&\dot{H}=f_{1}(R,H),\\
&\dot{R}=f_{2}(R,H).
\end{split}\end{equation}
The critical points are found by setting eqs.(\ref{26}) and
(\ref{27}) equal to zero. The results are
\begin{equation}\label{29}
\frac{1}{2}RF_{R}+GF_{G}-F=6b'H^{4},
\end{equation}
and
\begin{equation}\label{30}
R=12H^{2},
\end{equation}
from which eq.(\ref{22}) results in
\begin{equation}\label{31}
G=24H^{4}=\frac{R^{2}}{6}
\end{equation}
at critical points. The above equations must be solved to obtain
the coordinates of critical points in $R-H$ space, i.e. $(R_{c}$,
$H_{c})$. Note that in $b'\rightarrow 0$ limit, which is, in fact,
the $\hbar\rightarrow 0$ limit (mentioned after eq.(\ref{20})),
the critical point relations (\ref{29}) and (\ref{30}) are reduced
to classical relations obtained in~\cite{alaii}. It is also
interesting to note that from three quantum parameters $b$, $b'$
and $b''$, only the parameter $b'$ appears in critical point
equations (\ref{29}) and (\ref{30}).

To study the stability of each critical point, the eigenvalues of
the matrix
\begin{equation}\label{32}
M=\left(\begin{array}{ll}\partial{f_1}/\partial{H}&\partial{f_1}/\partial{R}\\
\partial{f_2}/\partial{H}&\partial{f_2}/\partial{R}
\end{array}\right)_{R=R_c,H=H_c}
\end{equation}
must be calculated. The index "$c$", denotes "critical value". The
critical point is a stable attractor if the real parts of all
eigenvalues are negative. The stable attractors divide to two
categories: it is "node" when the eigenvalues are real, and
"spiral" if they are complex conjugate. For real eigenvalues, but
with opposite sign, the critical point is a saddle point.

For autonomous eqs.(\ref{26}) and (\ref{27}), the matrix $M$
becomes
\begin{equation}\label{33}
M=\left(\begin{array}{cc} -4H&1/6\\
S&H
\end{array}\right)_{R=R_c,H=H_c}\: ,
\end{equation}
where
\begin{equation}\label{34}
S=\frac{-6F_{R}+(\frac{2}{3}b+b''-b')R}{3(F_{RR}+8H^{2}F_{RG}+16H^{4}F_{GG})-\frac{1}{2}(\frac{2}{3}b+b'')}.
\end{equation}
The eigenvalues of $M$ then become
\begin{equation}\label{35}
\lambda_{1,2}=\frac{1}{2}\left[-3H\pm\sqrt
{{(3H)}^{2}+\frac{2}{3}\left(S+24H^{2}\right)}\hspace{0.2cm}\right]_{R=R_{c},H=H_{c}}.
\end{equation}
To ensure a stable attractor, the real parts of $\lambda_{1}$ and
$\lambda_{2}$ must be negative, which satisfies if, and only if,
$S+24H^{2}<0$, or using eq.(\ref{30}):
\begin{equation}\label{36}
\delta=S+2R\biggm|_{R=R_{c},H=H_{c}}<0\,.
\end{equation}
At $b'\rightarrow 0$ limit, the above condition, multiplied by
$-1/{6}$, is reduced to
\begin{equation}\label{37}
\eta=\;\frac{F_{R}}{3(F_{RR}+8H^{2}F_{RG}+16H^{4}F_{GG})}-4H^{2}\biggm|_{R=R_{c},H=H_{c}}>0
\end{equation}
which is the same as the classical stability condition, obtained
in~\cite{alaii}. It must be noticed that in contrast to the
critical point equation (\ref{29}) which only depends on $b'$, the
stability condition (\ref{36}) depends on all three parameters
$b$, $b'$ and $b''$.

Before considering the complicated examples in the next section,
it is seen that for $R$-independent Lagrangian
\begin{equation}\label{38}
F(R,G)=F(G)
\end{equation}
where classically has no stable attractor, since
$\eta=-4H_{c}^{2}<0$, the quantum effects can induce the stable
attractors for these models. For example when $F_{GG}=0$,
eq.(\ref{36}) becomes
\begin{equation}\label{39}
{\delta}_{F(G)}=\,\frac{2b'}{\frac{2}{3}b+b''}R
\end{equation}
which is negative for the most background fields, i.e.
eq.(\ref{17}) shows that $b'<0$, except for
$28N_{\rm{HD}}>N+11N_{{1}/{2}}+62N{1}+1411N_{2}$. So many of the
$F(G)$ gravity models may have the stable quantum attractors.

\section{Some examples of $f(R)$ and $F(R,G)$ gravities}

In this section we study some specific examples to explore some
important features of quantum attractors.
\subsection{$F(R,G)=a_{1}+a_{2}G\ln G$ models}
As the first example, we consider a specific $F(G)$ gravity model
with action $a_{1}+a_{2}G\ln G$. The classical critical points are
found by eq.(\ref{29}), when $b'$ sets to zero,
\begin{equation}\label{40}
GF_{G}-F=0\hspace{2ex}\Rightarrow\hspace{2ex}
G_{c}^{\phantom{c}\rm{class.}}=\,\frac{a_{1}}{a_{2}}\hspace{1ex},
\end{equation}
or, using eq.(\ref{31}),
\begin{equation}\label{41}
R_{c}^{\phantom{c}\rm{class.}}=\sqrt{\frac{6a_{1}}{a_{2}}}\hspace{1ex}.
\end{equation}
At quantum level, eq.(\ref{29}) results in
\begin{equation}\label{42}
GF_{G}-F=\frac{b'}{4}
G\hspace{2ex}\Rightarrow\hspace{2ex}G_{c}^{\phantom{c}\rm{quant}.}=\,\frac{4
a_{1}}{4 a_{2}-b'}\hspace{1ex},
\end{equation}
from which
\begin{equation}\label{43}
R_{c}^{\phantom{c}\rm{quant}.}=\sqrt{\frac{24a_{1}}{4a_{2}-b'}}\hspace{1ex}.
\end{equation}
Depending on the parameters $a_{1}$, $a_{2}$ and $b'$, different
situations may arise. For example if $a_{1}a_{2}>0$, so that
$R_{c}^{\phantom{c}\rm{class.}}$ in eq.(\ref{41}) becomes a real
positive number, the classical critical point is an unstable
attractor (see after eq.(\ref{38})), but the quantum critical
point can be a stable attractor. In Figs.1-3, the results of the
numerical calculations of eqs.(\ref{26}) and (\ref{27}) for
$F(G)=1+2G\ln{G}$ model has been reported. Because of the
numerical factors of $b$ and $b'$ in eq.(\ref{17}), it is easier
to rescale $F(R,G)$s by $c$, where $c^{-1}=360{(4\pi)}^{2}$.
Therefore by $F=1+2G\ln{G}$, we mean
$F=\frac{1}{360{(4\pi)}^{2}}(1+2G\ln{G})$. Now we consider two
different background matters. We take
$N_{{1}/{2}}=N_{1}=N_{2}=N_{\mathrm{HD}}=0$. For $N=1$, one has
$b=3$ and $b'=-1$ and for $N=5$, $b=15$ and $b'=-5$. In Fig.1, It
can be seen that
$R_{c}^{\phantom{c}\rm{class.}}=\sqrt{6a_{1}/a_{2}}=\sqrt{3}$ is
not a stable attractor, Fig.2 shows that in $N=1$ case, the
quantum corrected attractor
$R_{c}^{\phantom{c}\rm{quant.}}=\sqrt{{24a_{1}}/{(4a_{2}-b')}}=\sqrt{24/9}$
is not yet stable, but in the case $N=5$, Fig.3 shows that
$R_{c}^{\phantom{c}\rm{quant.}}=\sqrt{24/13}$ is a stable
attractor.

\begin{figure}[h!]\label{Figure_1}
\centering
\includegraphics*[height=5cm,width=5cm]{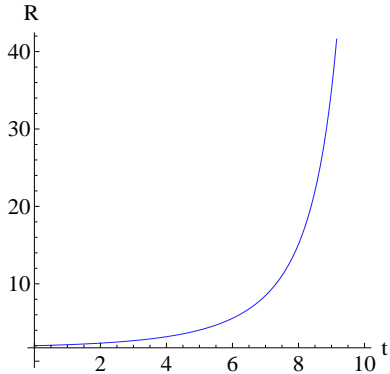}
\caption{ The plot of $R(t)$ of $F(R,G)=1+2G\ln{G}$ model at
classical level.}
\end{figure}
\begin{figure}\label{Figure_2}
\centering
\hspace{1cm}\includegraphics*[height=4cm,width=5cm]{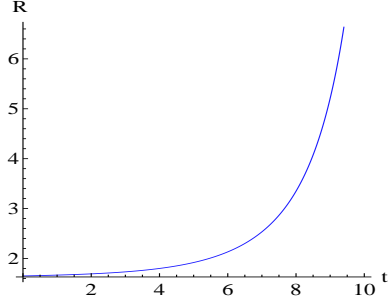}\hspace{1cm}
\caption{ The plot of $R(t)$ of $F(R,G)=1+2G\ln{G}$ model, with
$N=1$.}
\end{figure}
\begin{figure}\label{Figure 3}
\centering
\hspace{1cm}\includegraphics*[height=4cm,width=5cm]{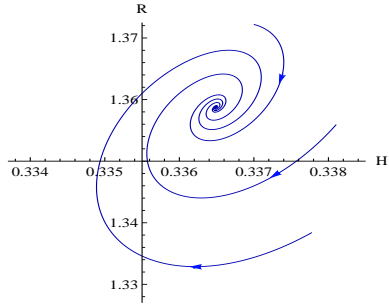}\hspace{1cm}
\caption{ The spiral paths in $R-H$ plane of $F(R,G)=1+2G\ln{G}$,
with $N=5$, at quantum level.}
\end{figure}

Other interesting cases also exist. For $a_{1}a_{2} <0$, where
classically there is no critical point, quantum effects can induce
a stable attractor. Fig.4 shows the phase-space of
$F(R,G)=1-\frac{1}{8}G\ln{G}$ model with $N=1$ background. As it
is seen, the point $(R_{c},H_{c})=\left(6.93,0.76\right)$ is a
stable attractor of this model, which has no analogous at
classical level.
\begin{figure}[hbt]\label{Figure_4}
\centering
\hspace{1cm}\includegraphics*[height=4cm,width=5cm]{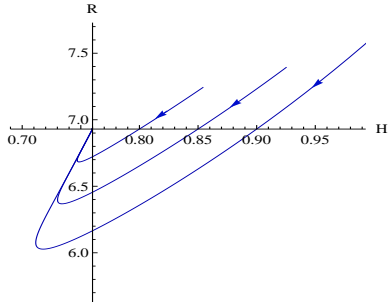}\hspace{1cm}
\caption{ The paths of $F(R,G)=1-\frac{1}{8}\ln{G}$, with $N=1$,
at quantum level.}
\end{figure}
\subsection{$F(R,G)=a_{1}+a_{2}R+a_{3}R^{2}$ models}

In the previous example, there is one critical point in each
classical and quantum mechanical levels, and at $b'\rightarrow 0$,
i.e. $\hbar\rightarrow 0$, limit,
$R_{c}^{\phantom{c}\rm{quant.}}\rightarrow
R_{c}^{\phantom{c}\rm{class.}}$. Now we consider the class of
models in which the number of classical and quantum mechanical
critical points are not the same, and more, at $b'\rightarrow 0$
limit, some of the quantum attractors do not lead to classical
ones.

Consider the following $f(R)$ gravity model
\begin{equation}\label{44}
f(R)=a_{1}+a_{2}R+a_{3}R^{2}\,.
\end{equation}
The only classical critical point is found by eq.(\ref{29}), with
$b'=0$, as following
\begin{equation}\label{45}
R_{c}^{\phantom{c}\rm{class.}}=-\frac{2a_{1}}{a_{2}}\,.
\end{equation}
The condition of stability of this solution, using eq.(\ref{37}),
is
\begin{equation}\label{46}
\eta=\frac{a_{2}}{6a_{3}}>0\,.
\end{equation}
So to have a positive curvature attractor, $a_{1}a_{2}$ must be
negative, and for its stability, $a_{2}a_{3}$ must be positive.
Note that because of eq.(\ref{30}), the negative values of $R_c$
are not acceptable, since they lead to imaginary Hubble constants,
which are not physical.

After adding the quantum correction terms, the action (\ref{44})
has two quantum critical points. Using (\ref{29}), one finds
\begin{equation}\begin{split}\label{47}
&R_{c1}^{\phantom{c1}\rm{quant.}}=\frac{-6a_{2}-\sqrt{{(6a_{2})}^{2}-24a_{1}b'}}{b'}\:,\\
&R_{c2}^{\phantom{c2}\rm{quant.}}=\frac{-6a_{2}+\sqrt{{(6a_{2})}^{2}-24a_{1}b'}}{b'}\:,
\end{split}\end{equation}
and their corresponding $\delta$s, which determine the stability
behavior of these points, are
\begin{equation}\label{48}
{\delta}_{1}\,=\,-\,{\delta}_{2}=\,\frac{2\sqrt{{(6a_{2})}^{2}-24a_{1}b'}}{12a_{3}-(\frac{2}{3}b+b'')}\,.
\end{equation}
Let us first study the $b'\rightarrow 0$ limit of curvatures in
eq.(\ref{47}), which their values depend on the sign of $a_{2}$:
\begin{equation}\label{49}
R_{c1}^{\phantom{c1}\rm{quant.}}\longrightarrow
-\frac{12a_{2}}{b'}
\hspace{2ex},\hspace{2ex}R_{c2}^{\phantom{c1}\rm{quant.}}\longrightarrow
-\frac{2a_{1}}{a_{2}}\,=R_{c}^{\phantom{c}\rm{class.}}\hspace{2ex},\hspace{2ex}\text{for}\hspace{2ex}a_{2}>0
\end{equation}
and
\begin{equation}\label{50}
R_{c1}^{\phantom{c1}\rm{quant.}}\longrightarrow
-\frac{2a_{1}}{a_{2}}=R_{c}^{\phantom{c}\rm{class.}}
\hspace{2ex},\hspace{2ex}R_{c2}^{\phantom{c2}\rm{quant.}}\longrightarrow
-\frac{12a_{2}}{b'}\,\hspace{2ex},\hspace{2ex}\text{for}\hspace{2ex}a_{2}<0.
\end{equation}
Now we encounter a new situation, one of the quantum curvatures
goes to classical value, and the other one diverges.

This behavior is the characteristic of "singular perturbation
theory". Mathematicians divide the perturbation theory to two
categories, "regular" and "singular". In regular perturbation
problem, the solution of the perturbed equation
\begin{equation}\label{51}
f_{1}(x)+\varepsilon f_{2}(x)=0
\end{equation}
is
\begin{equation}\label{52}
x({\varepsilon})=\sum_{n=0}^{\infty}\,a_{n}{\varepsilon}^{n}
\end{equation}
where $a_{0}$, i.e. the zero-order solution, is the solution of
the unperturbed equation:
\begin{equation}\label{53}
f_{1}(a_{0})=0\:.
\end{equation}
This is the usual behavior that we almost always encounter in
physics. But in some perturbation problems, the situation is
different. In singular perturbation problem, these two solutions,
i.e. the "zero-order solution" and the solution of "unperturbed
problem", do not coincide. In fact, the zero-order solution may
depend on $\varepsilon$ and may exist only for nonzero
$\varepsilon$. This situation occurs whenever the power of the
perturbation terms are greater than the unperturbed
terms~\cite{ben}. Consider, for example, the following algebraic
equation:
\begin{equation}\label{54}
\varepsilon x^{2}+x-1=0.
\end{equation}
Their solution are
\begin{equation}\label{55}
x=\frac{1}{2{\varepsilon}}\left(-1\pm\sqrt{1+4\varepsilon}\:\right)\hspace{2ex}\Rightarrow
\begin{cases}
x_{1}=1-{\varepsilon}+2{\varepsilon}^{2}+\cdots\\
x_{2}=-{1}/ {\varepsilon}-1+{\varepsilon}+\cdots
\end{cases}
\end{equation}
At $\varepsilon\rightarrow 0$ limit, $x_{1}\rightarrow 1$, which
is the solution of the unperturbed equation $x-1=0$, while
$x_{2}$, which diverges at $\varepsilon\rightarrow 0$, has no
unperturbed analogous.

This highly dependence of a problem to the perturbation is
frequently encountered in chaotic dynamical systems. The
appearance of this kind of solutions shows that one can not
discard the perturbative terms, i.e. it can not be ignored, sets
to zero, in the equations and therefore in the solutions.

For the example in hand, i.e. the Lagrangian (\ref{44}), the
critical point equation (\ref{29}) becomes
\begin{equation}\label{56}
-a_{1 }-\frac{1}{2}a_{2}R=\frac{b'}{24}R^{2},
\end{equation}
which is very similar to eq.(\ref{54}). So it is natural we
encounter the singular perturbation problem, with its mentioned
behaviors. In this problem, one classical solution, increases to
two quantum mechanical ones, which one of them, depending on the
sign of $a_{2}$, goes to classical solution, and the other one
blows up. This shows that one can not ignore the quantum
correction terms and they must be always added to classical
equations of motion. The classical attractor can be stable or not,
depending on the sign of $a_{2}a_{3}$ , see eq.(\ref{46}), but in
quantum case, one of the solutions is always stable and the other
is unstable, see eq.(\ref{48}).

We may have, depend on the parameters, one positive
$R_{c}^{\phantom{c}\rm{class.}}$ and one positive
$R_{c}^{\phantom{c}\rm{quant.}}$, no positive
$R_{c}^{\phantom{c}\rm{class.}}$ and one positive
$R_{c}^{\phantom{c}\rm{quant.}}$, or one positive
$R_{c}^{\phantom{c}\rm{class.}}$ and two positive
$R_{c}^{\phantom{c}\rm{quant.}}$. As a specific example for the
latter case, we consider the following action
\begin{equation}\label{57}
F(R)=-1+R-R^{2}
\end{equation}
in which we again write in $c^{-1}=360{(4\pi )}^{2}$ unit. Using
eqs.(\ref{45})-(\ref{48}), one finds
$R_{c}^{\phantom{c}\rm{class.}}=2$ with $\eta =-1/6$,
$R_{c1}^{\phantom{c}\rm{quant.}}=6+\sqrt{12}$ with
$\delta_1=-\sqrt{12}/7$, and
$R_{c2}^{\phantom{c1}\rm{quant.}}=6-\sqrt{12}$ with
$\delta_2=\sqrt{12}/7$. So $R_{c}^{\phantom{c}\rm{class.}}$ and
$R_{c2}^{\phantom{c1}\rm{quant.}}$ are unstable critical points
and tend together at $b'\rightarrow 0$ limit, while
$R_{c1}^{\phantom{c1}\rm{quant.}}$ is a stable attractor which
blows up at this limit, see eq.(\ref{49}). We have chosen $N=1$,
and other $N_i$'s equal to zero. Fig.5 shows the behavior of
$R(t)$ in classical regime, which verifies the unstability nature
of $R_{c}^{\phantom{c}\rm{class.}}$, and Fig.6 shows that
$R_{c1}^{\phantom{c1}\rm{quant.}}$ is a stable attractor.
\begin{figure}[hbt]\label{Figure_5}
\centering
\hspace{1cm}\includegraphics*[height=5cm,width=5cm]{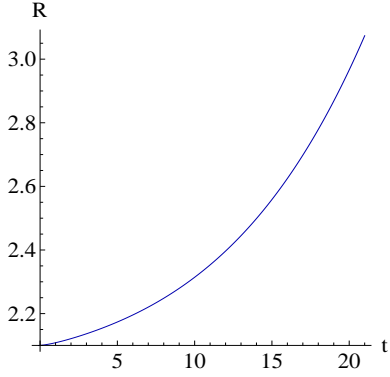}\hspace{1cm}
\caption{ The plot of $R(t)$ of $F(R)=-1+R-R^{2}$ model at
classical level.}
\end{figure}
\begin{figure}[hbt]\label{Figure_6}
\centering
\hspace{1cm}\includegraphics*[height=5cm,width=5cm]{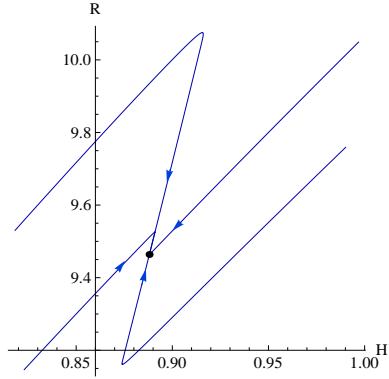}\hspace{1cm}
\caption{ The quantum attractor of $F(R)=-1+R-R^{2}$ model, when
 $N=1$. The stable attractor $R_{c1}=6+\sqrt{12}$ blows up at $b'\rightarrow 0$ limit.}
\end{figure}

 \subsection{$F(R,G)=a_{1}+a_{2}R+a_{3}R^{2}-a_{4}R^{2}\ln{R}$
 models}

 As other interesting example, we consider the following
 $f(R)$ action:
\begin{equation}\label{58}
f(R)=a_{1}+a_{2}R+a_{3}R^{2}-a_{4}R^{2}\ln{R}.
\end{equation}
This model has two classical and two quantum-mechanical critical
points, with coordinates
\begin{equation}\begin{split}\label{59}
&R_{c1}^{\phantom{c1}\rm{class.}}=\frac{-a_{2}-\sqrt{{(a_{2})}^{2}-8a_{1}a_{4}}}{2a_{4}}\,,\\
&R_{c2}^{\phantom{c2}\rm{class.}}=\frac{-a_{2}+\sqrt{{(a_{2})}^{2}-8a_{1}a_{4}}}{2a_{4}}\,,
\end{split}\end{equation}
and
\begin{equation}\begin{split}\label{60}
&R_{c1}^{\phantom{c1}\rm{quant.}}=\frac{-a_{2}-\sqrt{{(a_{2})}^{2}-8a_{1}(a_{4}+b'/12)}}{2a_{4}+{b'}/{6}}\,,\\
&R_{c2}^{\phantom{c2}\rm{quant.}}=\frac{-a_{2}+\sqrt{{(a_{2})}^{2}-8a_{1}(a_{4}+b'/12)}}{2a_{4}+{b'}/{6}}\,,
\end{split}\end{equation}
respectively. As $b'$ tends to zero, irrespective of other
parameters, it is clear from eqs.(\ref{59}) and (\ref{60}) that
\begin{equation}\label{61}
R_{ci}^{\phantom{ci}\rm{quant.}}\hspace{1ex}\xrightarrow{\hspace{2ex}b'\rightarrow
0\hspace{2ex}}\hspace{1ex}R_{ci}^{\phantom{ci}\rm{class.}}\hspace{3ex},\:i=1,2\,.
\end{equation}
However, this occurs whenever both classical and quantum critical
curvatures, separately for each $i$ in eq.(\ref{61}), belong to
the same region, i.e. positive or negative curvatures. But if they
are placed in different regions, e.g.
$R_{c1}^{\phantom{c1}\rm{class.}}<0$ and
$R_{c1}^{\phantom{c1}\rm{quant.}}>0$, we encounter a new problem.
As a specific example, if we demand
\begin{equation}\label{62}
R_{c1}^{\phantom{c1}\rm{class.}}<0\hspace{3ex},\hspace{3ex}R_{c2}^{\phantom{c2}\rm{class.}}>0
\hspace{3ex},\hspace{3ex}R_{c1}^{\phantom{c1}\rm{quant.}}>0
\hspace{3ex},\hspace{3ex}R_{c2}^{\phantom{c2}\rm{quant.}}>0,
\end{equation}
then the parameters must satisfy
\begin{equation}\label{63}
b'<0
\hspace{4ex},\hspace{4ex}a_{2}>0\hspace{4ex},\hspace{3ex}0<a_{4}<-\frac{b'}{12}\hspace{3ex},\hspace{3ex}
\frac{3{a_{2}}^{2}}{24a_{4}+2b'}\leq a_{1}<0.
\end{equation}
Because of the condition $0<a_{4}<-{b'}/{12}$, at $b'\rightarrow
0$ limit, the parameter $a_{4}$ must also goes to zero. So the
denominators of $R_{c1}^{\phantom{c1}\rm{class.}}$ and
$R_{c1}^{\phantom{c1}\rm{quant.}}$ both go to zero, with different
signs, while their numerators remain finite\footnote{Note that for
$R_{c2}^{\phantom{c2}\rm{class.}}$ and
$R_{c2}^{\phantom{c2}\rm{quant.}}$, also the denominators go to
zero, but at the same time, the numerators tend to zero, such that
$R_{c2}^{\phantom{c2}\rm{class.}}$ and
$R_{c2}^{\phantom{c2}\rm{quant.}}$ remain finite.}. Therefore
\begin{align}\label{64}
R_{c1}^{\phantom{c1}\rm{class.}}\hspace{.2cm}&\xrightarrow{\hspace{2ex}b'\rightarrow
0\hspace{2ex}}\hspace{2ex}-\infty
\nonumber\\
R_{c1}^{\phantom{c1}\rm{quant.}}\hspace{.1cm}&\xrightarrow{\hspace{2ex}b'\rightarrow
0\hspace{2ex}}\hspace{2ex}+\infty
\end{align}
This is another interesting behaviors of quantum $F(R,G)$ gravity
models. The quantum effects increase the number of positive
attractors, but the excess attractor blows up at $\hbar\rightarrow
0$ limit and does not approach to their corresponding classical
solution. As an explicit example belongs to this category, we
consider the following Lagrangian
\begin{equation}\label{65}
f(R)=-1+2R+\frac{5}{6}R^{2}-\frac{1}{6}R^{2}\ln{R}\;,
\end{equation}
and choose $N=3$ and other $N_{i}$'s are zero, so that $b'=-3$.
The critical points then become
\begin{align}\label{66}
\left(R_{c1}^{\phantom{c1}\rm{class.}},R_{c2}^{\phantom{c2}\rm{class.}}\right)
\hspace{.15cm}&=\left(\,-3(2+4/\sqrt{3})\,,\,3(-2+4/\sqrt{3})\,\right)
,
\nonumber\\
\left(R_{c1}^{\phantom{c1}\rm{quant.}},R_{c2}^{\phantom{c2}\rm{quant.}}\right)&=
\left(\,2(6+\sqrt{30})\,,\,2(6-\sqrt{30})\,\right).
\end{align}
So we have one acceptable (positive) classical and two quantum
mechanical critical points. All the critical points are stable
attractors. $R_{c2}^{\phantom{c2}\rm{quant.}}$ tends to
$R_{c2}^{\phantom{c2}\rm{class.}}$ at $b'\rightarrow 0$, while
$R_{c1}^{\phantom{c2}\rm{quant.}}$ has no classical analogous.
Figs.7 and 8 show the behaviors of the stable attractors of this
model at classical and quantum regimes, respectively.
\begin{figure}[hbt]\label{Figure_7}
\centering
\hspace{1cm}\includegraphics*[height=5cm,width=5cm]{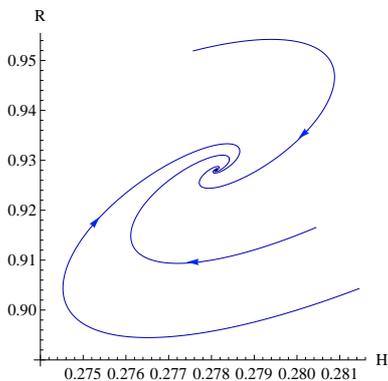}\hspace{1cm}
\caption{ The classical critical point of
$f(R)=-1+2R+\frac{5}{6}R^{2}-\frac{1}{6}R^{2}\ln{R}$ model.}
\end{figure}
\begin{figure}[hbt]\label{Figure_8}
\centering
\includegraphics*[height=5cm,width=4cm]{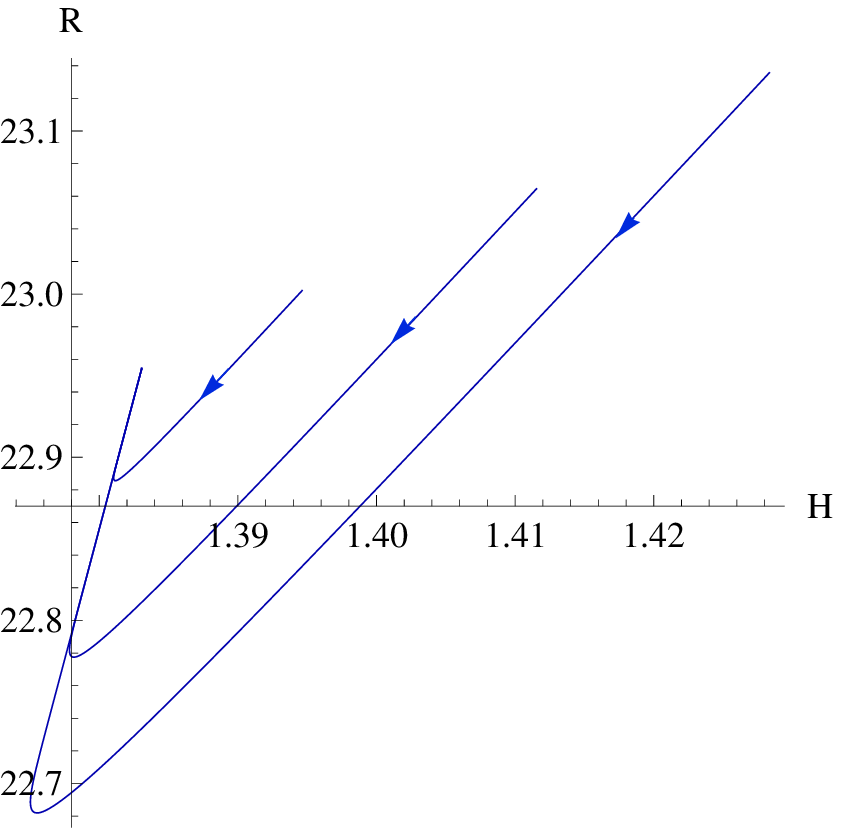}
\hspace{1cm}
\includegraphics*[height=5cm,width=4cm]{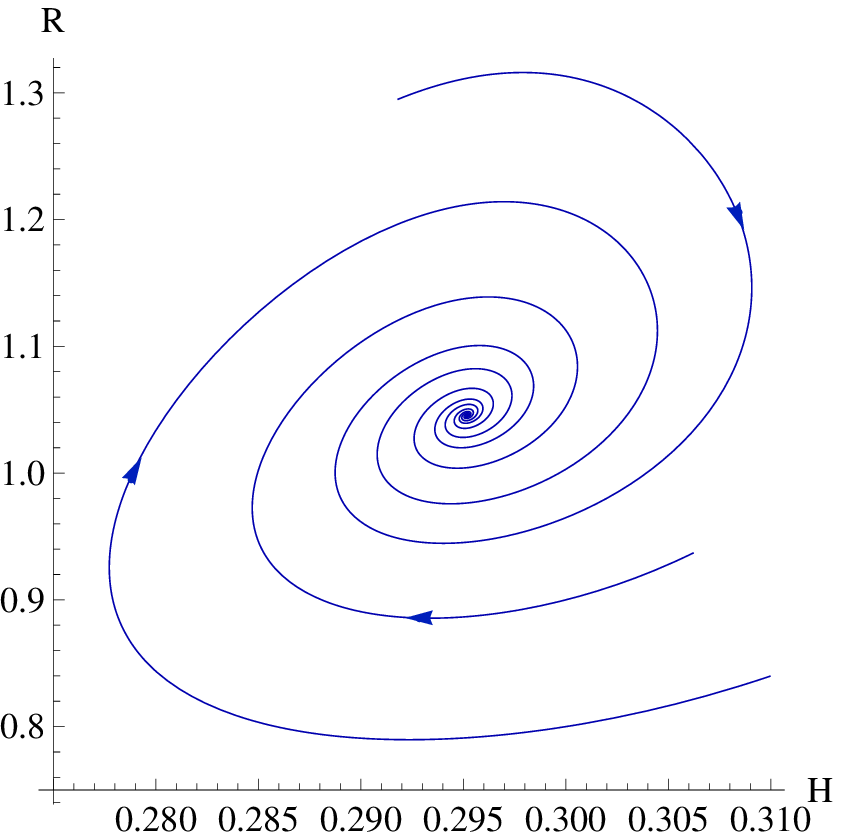}
\caption{ The quantum attractors of
$f(R)=-1+2R+\frac{5}{6}R^{2}-\frac{1}{6}R^{2}\ln{R}$ model, when
$N=3$.}
\end{figure}

The other interesting case of the Lagrangian (\ref{58}), is one in
which the classical curvatures become degenerate. This occurs when
$(a_{2})^{2}=8a_{1}a_{4}$. In this case, one has
\begin{equation}\label{67}
R_{c1}^{\phantom{c1}\rm{class.}}=R_{c2}^{\phantom{c2}\rm{class.}}=R_{\rm{class.}}=-\frac{a_{2}}{2a_{4}}\;,
\end{equation}
and
\begin{equation}\label{68}
R_{c1}^{\phantom{c1}\rm{quant.}}=\frac{R_{\rm{class.}}}{1-\sqrt{-\frac{b'}{12a_{4}}}}\hspace{2ex},\hspace{2ex}
R_{c2}^{\phantom{c2}\rm{quant.}}=\frac{R_{\rm{class.}}}{1+\sqrt{-\frac{b'}{12a_{4}}}}\:
.
\end{equation}
This phenomenon, in mathematics, is known as "bifurcation". In
dynamical systems, if the problem depends on one (or more)
parameter(s), and by continuous varying the parameter(s), two of
the critical points collide each other, it is said there is a
local bifurcation. In our problem, $b'$ plays the role of the
dynamical parameter. For $b'\neq 0$, we have two attractors
(\ref{68}). By decreasing $b'$ and reaching $b'=0$, they collide
and a single classical attractor (\ref{67}) is produced.

Another characteristic of this degenerate solution is its
$\eta$-value. To study the stability behaviors of
$R_{\rm{class.}}$, one must calculate $\eta$ in eq.(\ref{37}). For
$f(R)$ in eq.(\ref{58}), $\eta$ becomes, using eq.(\ref{30}),
\begin{equation}\label{69}
\eta=\frac{f_{R}}{3f_{RR}}-\frac{R}{3}=\frac{a_{2}+2a_{4}R}{3(2a_{3}-3a_{4}-2a_{4}\ln{R})}|_{R=R_{c}}
\end{equation}
So for $R_{\rm{class.}}$ in eq.(\ref{67}), one finds $\eta
|_{R=R_{\rm{class.}}=0}$. This means that in this case, the
eigenvalues of matrix $M$ are $\lambda_{1}=0$, $\lambda_{2}=-3H$
(see eq.(\ref{35}) for classical case, in which $S+24H^{2}$ must
be replaced by $-6\eta$). The appearance of zero eigenvalues in
dynamical systems is an important point, which one of its
consequences is that we can not predict the stability behavior of
the attractors. In this case, the first order variation, which
results in the stability matrix (\ref{32}), is not adequate and
one must consider the higher order approximations. This subject,
for $F(R,G)$ gravity models, has been discussed in~\cite{alaii}.

As an explicit example of the bifurcation point, we consider the
following example:
\begin{equation}\label{70}
f(R)=2-4R+2R^{2}-R^{2}\ln{R}
\end{equation}
and $N=1$. Then $R_{\rm{class.}}=2$ and
$R^{\rm{quant.}}_{\pm}=({24\pm2\sqrt{12}})/{11}$.
$R_{\rm{class.}}$ and $R^{\rm{quant.}}_+$ are unstable attractors,
while $R^{\rm{quant.}}_-$ is a stable attractor, see Fig.9.
\begin{figure}[h!]\label{Figure 9}
\centering
\hspace{1cm}\includegraphics*[height=5cm,width=5cm]{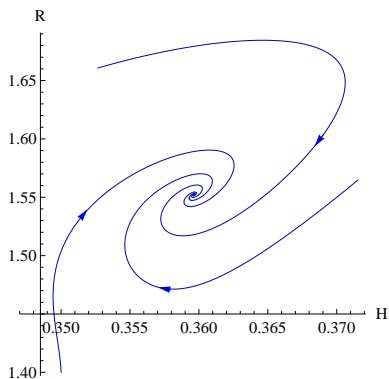}\hspace{1cm}
\caption{ The quantum critical point $R^{\rm{quant.}}_-$ of
$f(R)=2-4R+2R^{2}-R^{2}\ln{R}$ model with $N=1$.}
\end{figure}

As pointed out earlier, because of the degenerate nature of the
classical attractors of this example, one can not easily determine
the stability behavior of these critical points. In fact, if one
follows the method described in \cite{alaii}, it can be shown that
the next to leading order variation of Lagrangian (\ref{70}), near
$R_{\rm{class.}}=2$, is
\begin{equation}\label{n72}
\dot{\mathcal{U}}=6.9\,{\mathcal{U}}^2+\cdots,
\end{equation}
where the positive coefficient of ${\mathcal{U}}^2$-term, proves
the unstability of classical attractors near $R_{\rm{class.}}=2$.

\section{The matter field's contributions to attractors}
In this section we want to study the contribution of matter fields
to the structure of the phase-space, and investigate whether our
results in previous sections are changed.

By considering ${\rho}_{m}$, the autonomous system of equations
then consists of three equations, i.e. eqs.(\ref{26}), (\ref{27})
and eq.(\ref{14}), and therefore the phase-space becomes
3-dimensional, i.e. $(R,H,{\rho}_{m})$. It must be noted that the
variable $p_{m}$ in eq.(\ref{14}) can be expressed, by using the
equation of state of matter $p_{m}=f({\rho}_{m})$, in terms of
${\rho}_{m}$, so it is not an independent dynamical variable.

We restrict ourselves to barotropic fluid in which
\begin{equation}\label{71}
p_{m}={\omega}_{m}{\rho}_{m}\:,
\end{equation}
and ${\omega}_{m}$ is a constant. For example for dust,
${\omega}_{d}=0$, and for radiation, ${\omega}_{r}={1}/{3}$. In
this case, the third autonomous differential equation becomes
\begin{equation}\label{72}
\frac{d\rho_{m}}{dt}=-3H(1+{\omega}_{m}){\rho}_{m}\:.
\end{equation}
One can write the new system of autonomous equations as:
\begin{equation}\label{73}
\dot{H}=f_{1}(R,H,{\rho}_{m})\hspace{2ex},\hspace{2ex}\dot{R}=f_{2}(R,H,{\rho}_{m})\hspace{2ex},\hspace{2ex}
{\dot{\rho}}_{m}=f_{3}(R,H,{\rho}_{m})\:.
\end{equation}
The critical points are found by setting
$\dot{H}=\dot{R}={\dot{\rho}}_{m}=0$. In this way, the critical
point equations become the same as previous ones, i.e.
eqs.(\ref{29}) and (\ref{30}), and the following extra relation:
\begin{equation}\label{74}
{\rho}_{mc}=0.
\end{equation}
So the presence of the matter fields does not affect the position
of the attractors, which comes from eqs.(\ref{29}) and (\ref{30}).

The stability matrix $M$ now becomes a $3\times3$ matrix, as
follows
\begin{equation}\label{75}
M=\left(\begin{array}{ccc} -4H&1/6&0\\
S&H&B\\
0&0&-3(1+{\omega}_{m})H
\end{array}\right)_{R=R_{c},H=H_{c},{\rho}_{m}={\rho}_{mc}}\:,
\end{equation}
in which
\begin{equation}\label{76}
B=\frac{1}{6H(F_{RR}+8H^{2}F_{RG}+16H^{4}F_{GG})-(\frac{2}{3}b+b'')H}\bigg|_{R=R_{c},H=H_{c}}\:.
\end{equation}
The eigenvalues are: ${\lambda}_{1}$ and ${\lambda}_{2}$ in
eq.(\ref{35}), and
\begin{equation}\label{77}
{\lambda}_{3}=-3(1+{\omega}_{m})H_{c}\:.
\end{equation}
So, as long as
\begin{equation}\label{78}
{\omega}_{m}>-1\,,
\end{equation}
the eigenvalue ${\lambda}_{3}$ is negative and the stability
situation is completely determined by eq.(\ref{36}), or
(\ref{37}). In this way we obtain an important result: \emph{For
barotropic fluids with ${\omega}_{m}>-1$, the structure of the
phase-space is not changed.} Note that the condition (\ref{78})
satisfies by all ordinary, i.e. baryonic, matters.

The next point that we must consider is that the Lagrangian
(\ref{15}) has two kinds of effects in our problem. One of its
role is its quantum contribution to the $F(R,G)$ gravity, which we
have considered it by adding its $p_{A}$ and ${\rho}_{A}$ to the
Friedmann equations. The second role of (\ref{15}) is its
classical contribution to our problem, which we have not yet
considered it. To do so, one must consider the Friedmann equations
of combined actions (\ref{7}) and (\ref{15}), i.e.
\begin{equation}\label{79}
S=\int d^{4}x\sqrt{-g}{}\hspace{1ex}\left[F(R,G)+{\cal
L}_m+\frac{1}{2}\left(-\xi R\phi^2-(\nabla\phi)^2\right)\right].
\end{equation}
The other way is to consider the effects of extra action
(\ref{15}), by its energy density and pressure, i.e.
${\rho}_{\phi}$ and $p_{\phi}$, as the source terms for Friedmann
equations (\ref{26}) and (\ref{27}). This can be done by
considering the Friedmann equations of the scalar-tensor model
(\ref{15}) in the background of ordinary Einstein model, i.e.
\begin{equation}\label{80}
S=\frac{1}{2}\int d^{4}x\sqrt{-g}{}\hspace{1ex}\left[R -\xi
R\phi^2-(\nabla\phi)^2\right]\;,
\end{equation}
which are~\cite{alb}
\begin{equation}\label{81}
3H^2\,=\,{\rho}_{\phi}\,=\,3\xi
H^{2}{\phi}^2+\frac{1}{2}{\dot{\phi}}^{2}+6H\xi\phi{\dot{\phi}}\:,
\end{equation}
\begin{equation}\label{82}
-2\dot{H}\,=\,{\rho}_{\phi}+p_{\phi}\,=\,-2\xi\dot{H}{\phi}^{2}+2\xi
H\phi\dot{\phi}-2\xi\phi\ddot{\phi}+(1-2\xi){\dot{\phi}}^{2}\:,
\end{equation}
and
\begin{equation}\label{83}
\ddot{\phi}+3H\dot{\phi}+\xi\phi(6\dot{H}+12H^{2})\,=\,0\:.
\end{equation}
${\rho}_{\phi}$ is therefore given by eq.(\ref{81}). $p_{\phi}$
can be found by subtracting (\ref{81}) from (\ref{82}), using
(\ref{83}), which results in
\begin{equation}\label{84}
p_{\phi}=\left[2\xi (6\xi -1)\dot{H}+3\xi (8\xi
-1)H^{2}\right]{\phi}^{2}+2\xi H\phi\dot{\phi}+(\frac{1}{2}-2\xi
){\dot{\phi}}^{2}\:.
\end{equation}
Then the trace of energy-momentum tensor, $T_{\phi}$, becomes
\begin{equation}\label{85}
T_{\phi}=-{\rho}_{\phi}+3p_{\phi}=(1-6\xi ){\dot{\phi}}^{2}+6\xi
(6\xi -1)(\dot{H}+2H^{2}){\phi}^{2}.
\end{equation}
This is a complicated expression which becomes very simple at $\xi
={1}/{6}$. In fact
\begin{equation}\label{86}
T_{\phi}(\xi =\frac{1}{6})\,=\,0\:.
\end{equation}
This important equation has root in the conformal invariance of
the action (\ref{15}) at $\xi ={1}/{6}$, as pointed out after
eq.(\ref{15}).

Because of eq.(\ref{86}), we have
\begin{equation}\label{87}
-{\rho}_{\phi}+3p_{\phi}\,=\,0\hspace{2ex}\Rightarrow\hspace{2ex}p_{\phi}\,=\,\frac{1}{3}{\rho}_{\phi}\;.
\end{equation}
So one can add the classical contribution of action (\ref{15}) to
Friedmann equations by considering it as a barotropic fluid with
${\omega}_{\phi}={1}/{3}$! But we know that adding any barotropic
fluid with $\omega>-1$ does not change the structure of
phase-space, so this is also true for Lagrangian (\ref{15}). This
completes our proof.

\section{Conclusion}

In this paper we study the contribution of quantum phenomena on
the structure of $R-H$ phase-space of $F(R,G)$ gravity models. It
is shown that, both the location of de-sitter attractors and their
stability properties, can change because of quantum corrections.
These corrections are not, in general, from the type of small, and
therefore ignorable, corrections, instead the topology of
attractors may change. For the case where there is no attractor,
it may produce some attractors, an example of it is discussed in
section 3.1. In some cases, it increases the number of the
attractors and their stability characteristics, with an important
property: The increased attractors do not converge to classical
ones at $\hbar\rightarrow 0$ limit, and more, their locations blow
up. A fact which is known in the context of "singular
perturbation". An example of this kind of behaviors is given in
section 3.2. Also it is possible that the quantum contributions
change the attractors, such that not only they do not converge to
classical locations, but also the quantum and classical attractors
go more distant from each other at $\hbar\rightarrow 0$ limit.

It is also shown that the barotropic perfect fluids with $\omega
>-1$, do not change our results. Interestingly, the
coupled-quintessence Lagrangian, which is responsible for
producing the quantum terms, is from this type.

Besides the attractors studied in this paper, there are two other
classes of attractors in $F(R,G)$ gravity models~\cite{alaii}. One
of them is the singularities of $F(R,G)$ Lagrangian, which is
always lead to the stable attractors, and the other class is one
known as critical curve. The curve $R=12H^{2}$ is the location of
infinite number of the stable attractors for the models whose
Lagrangians satisfy eq.(\ref{29}) (with $b'= 0$). That is,
eq.(\ref{29}) (with $b'= 0$) satisfies for all $R$ and $H$s.
$F(R,G)=R^{2}g(G/R^{2}), F(R,G)=\alpha G+\beta R^{2},
F(R,G)=R-6G/R, \cdots $, are examples of these
Lagrangians~\cite{alaii}. Now it can be easily shown that the, so
called, singular attractors do not change because of the quantum
terms ${\rho}_{A}$ and $p_{A}$, but the critical curve
$R=12H^{2}$, reduced to a single point. The reason is easy. If
$F(R,G)$ satisfies
\begin{equation}\label{88}
\frac{1}{2}RF_{R}+GF_{G}-F=0
\end{equation}
for all $R$, then for these actions, the equation of critical
point (\ref{29}) becomes
\begin{equation}\label{89}
0=6b'H^{4}=\frac{1}{24}b'R^{2}\:,
\end{equation}
which has a unique solution $R_{c}=0$. So the quantum effects
reduce the infinite number of stable attractor to an attractor at
$R_{c}=0$.

Of course, the critical curve $R=12H^{2}$ still exists in our
case, but for $F(R,G)$s which satisfy
\begin{equation}\label{90}
\frac{1}{2}RF_{R}+GF_{G}-F=\frac{1}{24}b'R^{2}
\end{equation}
for all $R$ and $H$s. For these functions, since eq.(\ref{29})
satisfies by $F(R,G)$, this equation does not impose any
constraint on $R$ and $H$, and eq.(\ref{30}) is the only equation
which specifies the critical points. So $R=12H^{2}$ becomes the
critical curve of these models. An example of these kinds of
models is
\begin{equation}\label{91}
f(R)=(C+\frac{b'}{12}\ln{R})R^{2}\,,
\end{equation}
which satisfies eq.(\ref{90}). At $b'\rightarrow 0$ limit, it
gives $f(R)=R^{2}$, which is stated before (the $F(R,G)=\alpha
G+\beta R^{2}$ model, with $\alpha =0$).

{\bf Acknowledgement:} This work was partially supported by the
"center of excellence in structure of matter" of the Department of
Physics of the University of Tehran, and also a research grant
from the University of Tehran.\\ \\

\end{document}